# Automatic segmentation of novel coronavirus pneumonia lesions in CT images utilizing deep-supervised ensemble learning network

**Yuanyuan Peng[1,2*], Zixu Zhang[1], Hongbin Tu[1,3], Xiong Li[4]**

[1]School of Electrical and Automation Engineering, East China Jiaotong University, Nanchang, China

[2]School of Computer Science, Northwestern Polytechnical University, Xi'an, China

[3]Technique Center, Hunan Great Wall Technology Information Company Limited, Changsha, China

[4]School of software, East China Jiaotong University, Nanchang, China

**\* Correspondence:**
Yuanyuan Peng
3066@ecjtu.edu.cn

**Background:** The 2019 novel coronavirus disease (COVID-19) has been spread widely in the world, causing a huge threat to people's living environment.

**Objective:** Under computed tomography (CT) imaging, the structure features of COVID-19 lesions are complicated and varied greatly in different cases. To accurately locate COVID-19 lesions and assist doctors to make the best diagnosis and treatment plan, a deep-supervised ensemble learning network is presented for COVID-19 lesion segmentation in CT images.

**Methods:** Considering the fact that a large number of COVID-19 CT images and the corresponding lesion annotations are difficult to obtained, a transfer learning strategy is employed to make up for the shortcoming and alleviate the overfitting problem. Based on the reality that traditional single deep learning framework is difficult to extract complicated and varied COVID-19 lesion features effectively, which may cause some lesions to be undetected. To overcome the problem, a deep-supervised ensemble learning network is presented to combine with local and global features for COVID-19 lesion segmentation.

**Results:** The performance of the proposed method was validated in experiments with a publicly available dataset. Compared with manual annotations, the proposed method acquired a high intersection over union (IoU) of 0.7279 and a low hausdorff distance (H) of 92.4604.

**Conclusion:** A deep-supervised ensemble learning network was presented for coronavirus pneumonia lesion segmentation in CT images. The effectiveness of the proposed method was verified by visual inspection and quantitative evaluation. Experimental results indicated that the proposed method has a good performance in COVID-19 lesion segmentation.

## 1 INTRODUCTION

Since the end of 2019, acute infectious pneumonia characterized by novel coronavirus disease (COVID-19) infection has been rapidly spread in the world, posing a huge threat to people's lives [1]. The outbreak of the pneumonia caused by COVID-19 infection has been identified by the World Health Organization as an emergency public health event of international concern, the number of COVID-19 patients is rapidly growing in the world [2]. So far, the cumulative confirmed cases of COVID-19 in the world exceeded 200 million, and the cumulative deaths reached 3.6 million. The early symptoms of the pneumonia are not obvious, but are strongly infective. It has been caused huge economic losses to society and aroused wide concern in the world [3, 4].

The difficulty of prevention and treatment of COVID-19 has put forward urgent requirements to the research of rapid diagnosis methods. The prevention and control measures of early diagnosis, early isolation and early treatment for COVID-19 patients are one of the most effective strategies to solve the pneumonia epidemic [5]. However, widely used nucleic acid testing and specific antibody detection technologies have several disadvantages such as lagging, long time consuming, low detection efficiency, and serious risk of missed detection [6]. As one of the most effective lung imaging mode, computed tomography (CT) has the ability to identify the changes in lung lesions and pathological features in patients with COVID-19 [7,8]. Therefore, a large number of researchers have designed many different deep learning models to assist clinic doctors in the rapid diagnosis of COVID-19 in CT images [9].

Table I. COVID-19 lesion segmentation with different methods.

| | Methods | Advantages | disadvantages |
|---|---|---|---|
| Deep-supervised learning | Yazdekhasty et al.[18] Wang et al.[19] Gao et al.[20] | Good performance in sufficient data and model generalization | Mounts of voxel-based Data annotation |
| Semi-supervised learning | Zhao et al.[21] Abdel-Basset et al.[22] | Good performance in lacking data | Parts of voxel-based Data annotation |
| Weakly supervised learning | Yang et al.[25] Laradji et al.[26] Wu et al.[27] Wang et al.[28] | Class annotation | Poor segmentation in small lesions |
| Unsupervised learning | Yao et al.[29] | No data annotations | Bad performance |

Traditional methods achieve the purpose of different semantic segmentation tasks by extracting features of the target objects [10-12], but the segmentation performance wasn't good enough. To overcome the problem, various of deep learning frameworks have been proposed to effectively segment target objects [13]. It could be divided into four classifications, deep-supervised learning, semi-supervised learning, weakly supervised learning and unsupervised learning approaches in COVID-19 lesion segmentation. Compared with traditional methods, the segmentation performance has been largely improved by deep learning networks [14-16].

Although deep learning frameworks can achieve a good performance in COVID-19 lesion segmentation, it required a lot of data and the corresponding data annotations [17]. To solve the problem, data augmentation is one of the most common operations, which can generate large mounts of data through rotation, scaling, clipping, and transpose. However, using only data augmentation may cause some lesions to be undetected. Therefore, many advanced strategies such as transfer learning, multi-task approach and attention mechanism have been proposed to improve the performance of COVID-19 lesion segmentation. Based on this theory, a multi-task approach was designed by Yazdekhasty et al. to reach the purpose [18], it had a good performance in lacking data and model generalization. Using a different strategy, Wang et al. integrated with transfer learning, unet model and multi-task learning to improve the segmentation performance of COVID-19 lesions [19]. Recently, attention mechanism [19, 20] was employed to make up for the shortcoming of partial information missed caused by convolution operation in deep learning networks. In order to further improve the segmentation performance, semi-supervised learning strategies were proposed to train mounts of pseudo annotations. Based on this strategy, Zhao et al. presented a randomly selected propagation strategy to improve the segmentation performance of COVID-19 lesions [21]. Similarly,

Abdel-Basset et al. proposed an innovative semi-supervised few-shot segmentation method for COVID-19 lesion segmentation from a few amounts of annotated lung CT images [22]. Existing supervised and semi-supervised methods require mounts of voxel-based annotations in training stage [23]. Unfortunately, it is difficult for clinicians to precisely annotate COVID-19 lesions due to the complex structural changes and blurred boundary information [24]. To overcome the problem, mounts of weakly supervised methods have been proposed to segment COVID-19 lesions. The advantage of the weakly supervised approaches is that it can replace the complicated COVID-19 lesion labels with simple ones for training. Based on this strategy, Yang et al. presented a weakly supervised method based on generative adversarial network (GAN) to improve the accuracy of COVID-19 lesion segmentation [25]. A generator was adopted to remove lesions and generate healthy slices from input images, while a discriminator was used to force the generator to generate more accurate results with mounts of image-level annotations. The method was improved by Laradji et al. [26], where they utilized two encoder-decoder frameworks with shared weights. The first one encoded the original images, and the point annotations were treated as the corresponding supervised term. While the second one encoded the original image with geometric transformation, and the outputs of the first one with geometric transformation were regarded as the corresponding supervised term. Similarly, Wu et al. proposed a new 3D active learning framework called COVID-AL to segment COVID-19 lesions with volume-annotations [27]. Recently, Wang et al. proposed a weakly supervised deep learning framework for COVID-19 lesion segmentation [28]. Firstly, a pre-trained unet is applied to remove unrelated tissues for lung segmentation. Subsequently, the segmented lung is fed into the designed DeCoVNet to acquire the COVID-19 lesion feature map. Finally, a class activation mapping algorithm and a 3D connected component algorithm were combined for COVID-19 lesion localization. The fatal flaw of the deep supervised approach, semi-supervised approach and weakly supervised approach is that mounts of data labels are required to supervise the training model in training stage. Whereas data annotations need clinical experts to spend a lot of time to annotate it. Different from above deep learning approaches, unsupervised approach has a good performance in objects segmentation without any annotated labels. To alleviate the burden of data annotation, Yao et al. designed an unsupervised NormNet model to distinguish COVID-19 lesions from complex lung tissues [29]. Based on the observation that parts of tracheae and vessels exhibit strong patterns, a three stage (random shape, nosie generation and image filtering) strategy was used to generate lesion shape for subsequent segmentation. Taking the difference between COVID-19 lesions and other tissues into consideration, a novel method named NormNet based on generative adversarial networks was presented for COVID-19 lesion segmentation. Unfortunately, the unsupervised NormNet model had a bad performance than some supervised methods. The advantages and disadvantages of different types of deep learning methods are summarized in Table I.

Motivated by the fact that different deep learning methods have their own unique advantages, these advantages can be fused using a deep-supervised ensemble learning network to improve the COVID-19 lesion segmentation results in CT images. Unfortunately, the specific deep learning framwork may take up more time and space.

In this study, a deep-supervised ensemble learning network is proposed for COVID-19 lesion segmentation in CT images. To alleviate the overfitting problem on small datasets, a transfer learning strategy is used to acquire initialization parameters with better feature performance. Subsequently, an enumeration grid model is exploited to estimate the optimal weights for multiple deep learning model integration. In particular, we pay special attention to COVID-19 lesion and its boundary segmentation, which can illustate the effectiveness of the proposed method. Compared with several methods, the proposed model has a good performance in COVID-19 lesion segmentation in CT images.

## 2    MATERIALS AND METHODS

In this study, we present a deep-supervised ensemble learning network as an alternative model to segment COVID-19 lesions comparing to several models in publicly available datasets. Visual inspection and quantitative evaluation were established in this study to verify the proposed ensemble learning network.

### 2.1    Evaluation criteria

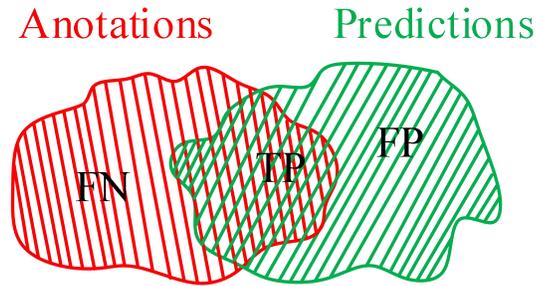

**FIGURE 1**| IoU criterion.

To illustrate the validation of the proposed method, an IoU criterion and $F_1$-measure [10] are applied to verify the good performance. As shown in **Figure 1**, annotations A are divided into false negative (FN) and true positive (TP), whereas predictions B are divided into TP and false positive (FP). In which, TP is the common region between annotations A and predictions B. Therefore, the mathematical description of IoU, Precison (P), Recall (R) and $F_1$ can be defined as following:

$$IoU = \frac{TP}{TP + FP + FN} \tag{1}$$

$$P = \frac{TP}{TP + FP} \tag{2}$$

$$R = \frac{TP}{TP + FN} \tag{3}$$

$$F_1 = \frac{2 \times P \times R}{P + R} \tag{4}$$

In other words, the greater proportion of the common region between annotations and predictions, the greater IoU and $F_1$ values will be. The smaller proportion of common region between annotations and predictions, the smaller IoU and $F_1$ values will be. Especially in the case with small COVID-19 lesions, IoU has a strong ability to illustrate the effectiveness of the proposed method. Additionally, $F_1$-measure criterion denotes the smilarity between annotations and predictions.

As we all know that IoU and $F_1$ pay more attention to the regional sensitivity of image segmentation, but the description of the segmentation boundary is also important. In order to further illustrate the effectiveness of the proposed method, hausdorff distance is employed to evaluate the validation of the deep learning framework. The hausdorff distance from annotations A to predictions B can be defined as

$$h(A,B) = \max_{a \in A} \left\{ \min_{b \in B} \left\{ d(a,b) \right\} \right\} \tag{5}$$

Where $h(A,B)$ is the hausdorff distance from A to B, $d(a,b)$ is the distance between point a and point b. In which, point a and b are belonging to A and B, respectively. To illustrate the distance between A and B, a more general definition would be :

$$H(A,B) = \max \left\{ h(A,B), h(B,A) \right\} \tag{6}$$

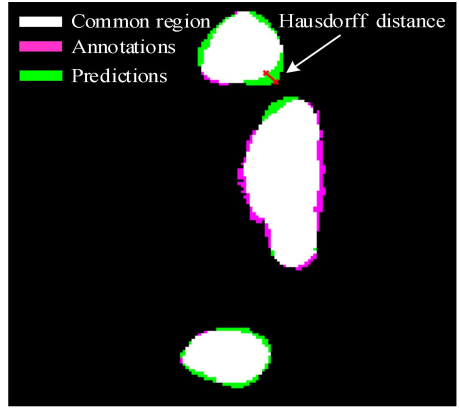

**FIGURE 2|** Hausdorff distance.

As shown in **Figure 2**, the pink area is the annotations A, the green area is the predictions B, and the white area is the common region covered by both predictions A and annotations B. Hausdorff distance measures the biggest distance from both predictions A and annotations B to common region. In order to avoid the influence of noise on Hausdorff distance, 95% hausdorff distance (HD95) are treated as useful values sorted from small to large. To avoid hausdorff distance being too large, a simple mathematical transformation is given

$$H = \sum_{i=1}^{n} \lg(HD95_i + 1) \tag{7}$$

Where $i$ is the number of the test CT slices. In other words, the better detection of the COVID-19 lesion boundary, the smaller $H$ value will be. Whereas the worse detection of the COVID-19 lesion boundary, the larger $H$ value will be. The evaluation criteria IoU, $F_1$ and $H$ are used to evaluate the validation of the proposed method.

## 2.2 Data and annotations

We employed a dataset from the China consortium of chest CT image investigation[30], which is a publicly available dataset. The dataset includes 150 CT scans. In which, 750 CT slices were selected from the dataset and annotated by four doctors with extensive clinical experience, 400 CT slices were treated as training set, 200 CT slices were considered as validation set, and 150 CT slices were used as test set. In this work, we treated the corresponding annotations as the groundtruth.

## 2.3 Overview of the proposed method

In this paper, we present a deep-supervised ensemble learning network for COVID-19 lesion segmentation in CT images in **Figure 3**. Firstly, data augmentation is applied to increase the training data and improve the generalization ability of the model. Subsequently, a transfer learning strategy is employed to copy with small datasets and alleviate the overfitting problem. Finally, a deep-supervised ensemble learning network is presented to combine with local and global features for COVID-19 lesion segmentation in CT images.

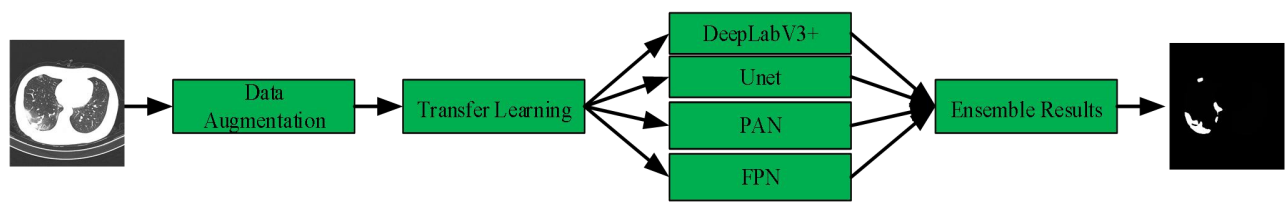

**FIGURE 3|** A pipeline for COVID-19 lesion segmentation in CT images.

## 2.4    Data augmentation

Clinicians spend a great deal of time to annotate complex structures of COVID-19 lesions, which is too expensive. In general, data augmentation [31] is applied to process and increase the training data, so as to make the data as diverse as possible and make the training model generalization ability stronger. In order to reach the purpose, CT image operations such as horizontal flip, vertical flip, rotation and scaling are adopted to increase training data and the corresponding COVID-19 lesion annotations. After data augmentation, the number of training data is varied from 400 to 4000.

## 2.5    Transfer learning

Deep learning models have been widely applied in medical image processing [32]. In order to acquire accurate COVID-19 lesion segmentation results, a large amount of CT images and manual annotations are required to adjust the parameters for the special deep learning model. However, it is too difficult to obtain mounts of CT images and the corresponding manual lesion annotations. In general, the designed deep learning models are trained with small datasets, which may leads to poor generalization ability and serious overfitting performance. To copy with the shortcomings, many technical means such as data augmentation [31], multi-task learning [33], transfer learning [34] and attention mechanism [35] can be used to achieve the segmentation task. In this section, a transfer learning strategy is applied to solve the shortcomings.

It is a challenging task for doctors to manually annotate complex and variable COVID-19 lesions. In order to acquire the best parameters of the designed model with the limited COVID-19 lesion annotations, a transfer learning strategy is employed to accurately segment COVID-19 lesions in CT images. In which, ImageNet dataset is treated as the pre-training dataset, which is one of the largest image datasets in the world [36]. Whereas the EfficientNet model [37, 38] is considered as the pre-training model. First, the model is trained with the training data and the corresponding manual annotations to generate pseudo annotations, it was treated as the teacher in the "teacher-student" model. Whereas a powerful EfficientNet model is retrained by using manual and pseudo annotations, it was considered as the student in the "teacher-student" model. In the student learning stage, adding noise processing is used to make students′ generalization ability better than teachers′ [39]. In this section, the estimated parameters are regarded as the initialized parameters for the deep-supervised ensemble learning network.

## 2.6    Deep-supervised ensemble learning  network

Different networks have different advantages and disadvantages, the advantages can be integrated together by effectively integrating various networks. To acquire the best COVID-19 lesions segmentation performance, a deep-supervised ensemble learning network is presented for COVID-19 segmentation in CT images.

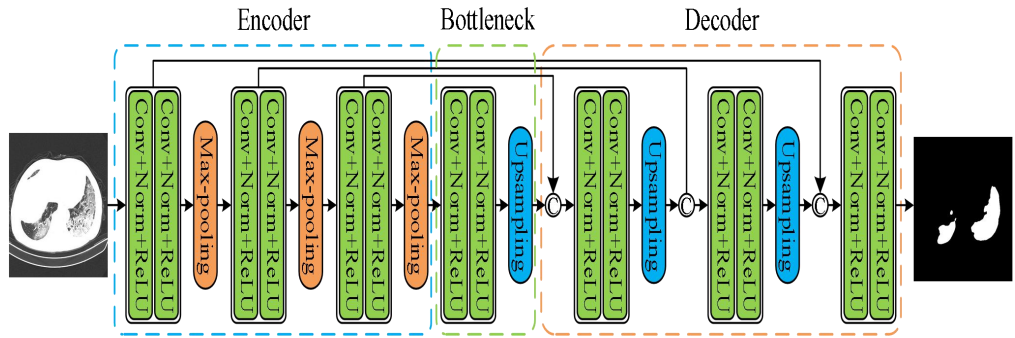

**FIGURE 4|** Unet model.

The Unet model was applied according to Su et al. and Wang et al. [40, 41]. As shown in **Figure 4**, the network had three encoding and decoding blocks, respectively. Encoding was designed using max-pooling, whereas the decoding was performed via a deconvolution. In addition, the encoder and the decoder were connected via skip connections.

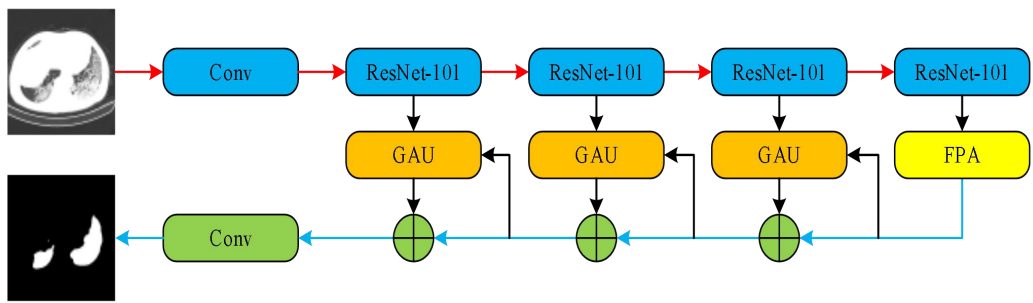

**FIGURE 5|** Pyramid attention network.

As a second architecture we implemented the pyramid attention network (PAN) by Li et al.[42]. Unlike the traditional deep learning model, the basic principle of the pyramid attention network is to effectively extract local and global features of target objects by integrating attention mechanism and spatial pyramid structures. As shown in **Figure 5**, an encoder-decoder scheme was adopted to locate target objects. In the encoder module, a feature pyramid attention (FPA) was introduced to adopted spatial pyramid attention mechanism in the high-level output, and the global information was applied to learn stronger feature representation. In the decoder module, a global attention upsample (GAU) module was applied to extract the global information of the target objects, so as to effectively segment COVID-19 lesions.

The third implemented architecture is the DeepLabv3+ by Chen et al. [43]. As shown in **Figure 6**, it is the DeepLabv3 extended version by adding a decoder module to refine the segmentation results especially along the novel coronavirus pneumonia lesion boundaries. Specifically, the Xception model was explored and the depthwise separable convolution was applied to both atrous spatial pyramid pooling and decoder modules, resulting in a faster and stronger encoder-decoder network.

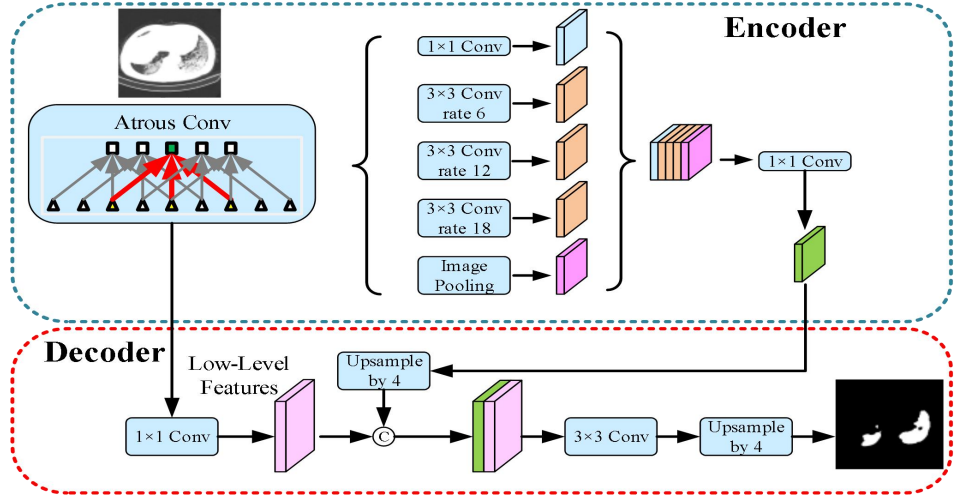

**FIGURE 6 |** DeepLabv3+ model.

The fourth implemented architecture is the multi-scale feature pyramid network (FPN) by Lin et al. [44]. As shown in **Figure 7**, a top-down architecture with skip connections was designed to express high-level feature maps at all scales. Subsequently, The semantic feature maps with different scales were intergrated to improve the segmentation performance of COVID-19.

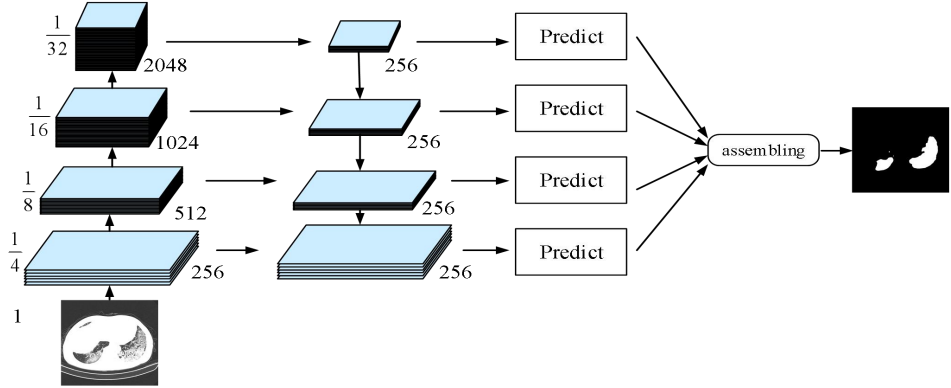

**FIGURE 7 |** FPN model.

In these deep learning networks, training was performed using the Adam optimizer with a learning rate of $10^{-5}$, the dice loss was considered as the loss function to express the relationship between the predicted probabilities and the corresponding lesion annotations. Inspired by the previous work of Golla et al.[45], we present an ensemble module to ensemble the probability feature maps of four networks. To give a mathematical express:

$$E = w_1 \times E_1 + w_2 \times E_2 + w_3 \times E_3 + w_4 \times E_4 \tag{8}$$

Where $w_1$, $w_2$, $w_3$ and $w_4$ are the wighting parameters, $E_1$, $E_2$, $E_3$ and $E_4$ are represent the predicted probabilities of PAN, FPN, Unet, and Deeplabv3+ networks. In which, the relationship among the wighting parameters $w_1$, $w_2$, $w_3$ and $w_4$ can be represented as following

$$w_1 + w_2 + w_3 + w_4 = 1 \tag{9}$$

In order to acquire the most effective segmentation results of COVID-19 lesions in CT images, an enumeration grid model [46] is used to achieve the optimal weighting parameters.

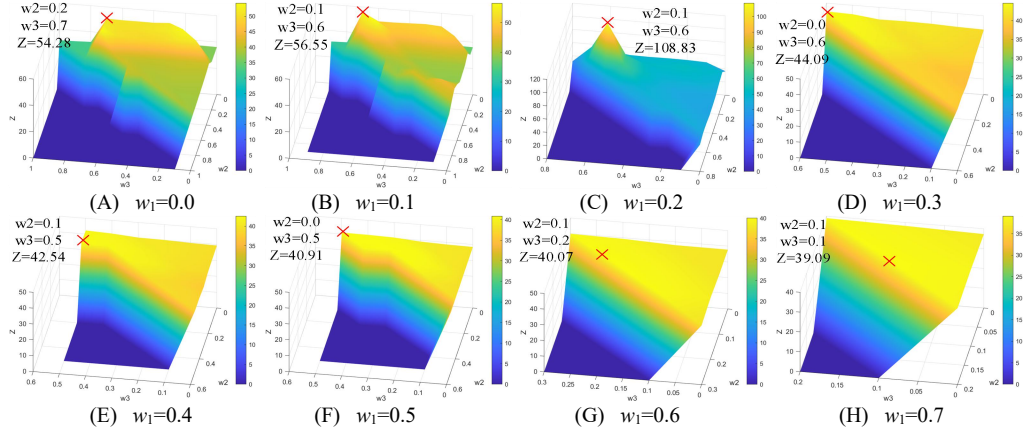

**FIGURE 8|** Weighting parameters optimization. (A):$w_1$ =0.0. (B):$w_1$ =0.1. (C):$w_1$ =0.2. (D):$w_1$ =0.3. (E):$w_1$ =0.4. (F):$w_1$ =0.5. (G):$w_1$ =0.6. (H):$w_1$ =0.7.

Weighting parameters play an important role in COVID-19 lesion segmentation. To overcome the problem, a simple but effective approach is designed to acquire the optimal weighting parameters with the enumeration grid model. Which is a traversal method, it has the ability to enumerate all the parameters $w_1$ , $w_2$ , $w_3$ and $w_4$ . Firstly, $w_1$ is set to be a fixed value varied from 0.0 to 1.0, $w_2$ and $w_3$ are treated as variable values. As we all known that the sum of all the weighting parameters is 1.0, the weighting parameter $w_4$ is $1-w_2-w_3$ . In other words, $w_4$ is defined by variable $w_2$ and $w_3$ . As a result, all of the wighting parameters and the corresponding quantitative index intersection over union (IOU) can be acquired with the enumeration grid model. Because of the IoU value difference is very small, it is difficult to distinguish. To better distinguish the quantitative index IoU for a good presentation, a simple mathematical transformation is given

$$Z = \log_{9/10}^{|IoU-\max(IoU)-0.0001|}$$ (10)

As shown in **Figure 8**, the maximum values Z are marked with red cross and the corresponding values are described in **Table II**. It can be seen that the optimal weighting parameter are 0.2, 0.1, 0.6 and 0.1.

Table II. The maximum values Z

| Maximum points | | | | Maximum values |
|---|---|---|---|---|
| $w_1$ | $w_2$ | $w_3$ | $w_4$ | Z |
| 0.0 | 0.2 | 0.7 | 0.1 | 54.28 |
| 0.1 | 0.1 | 0.6 | 0.2 | 56.55 |
| 0.2 | 0.1 | 0.6 | 0.1 | **108.83** |
| 0.3 | 0.0 | 0.6 | 0.1 | 44.09 |
| 0.4 | 0.1 | 0.5 | 0 | 42.54 |
| 0.5 | 0.0 | 0.5 | 0 | 40.91 |
| 0.6 | 0.1 | 0.2 | 0.1 | 40.07 |
| 0.7 | 0.1 | 0.1 | 0.1 | 39.09 |
| 0.8 | 0 | 0.1 | 0.1 | 37.40 |
| 0.9 | 0 | 0 | 0.1 | 35.81 |
| 1.0 | 0 | 0 | 0 | 35.08 |

**Table III** lists the architecture parameters with different networks. The dice loss function was regarded as the loss function. Adaptive moment estimation (Adam) was employed for the training process, which iteratively updates different network weights based on a publicly available novel coronavirus pneumonia dataset. The learning rate was initialized with 10-5. The above deep learning model was implemented in python using pytorch with Lenovo Ren-9000 34IMZ, GPU GFX 2060, CPU 32G.

Table III.Architecture parameters with different networks.

| Parameters | Value |
|---|---|
| Input image size | 512×512 |
| Output image size | 512×512 |
| Learning rate | $10^{-5}$ |
| Activation layers | Adam |
| Epochs | 300 |
| Batch size | 4 |
| Loss function | Dice |

## 3 RESULTS

We present a deep-supervised ensemble learning network as an alternative model to segment COVID-19 lesions in CT images. The segmentation performance of the proposed method is validated in experiments with a publicly available dataset. The effectiveness of the proposed method was verified by visual inspection and quantitative evaluation.

### 3.1 Visual inspection

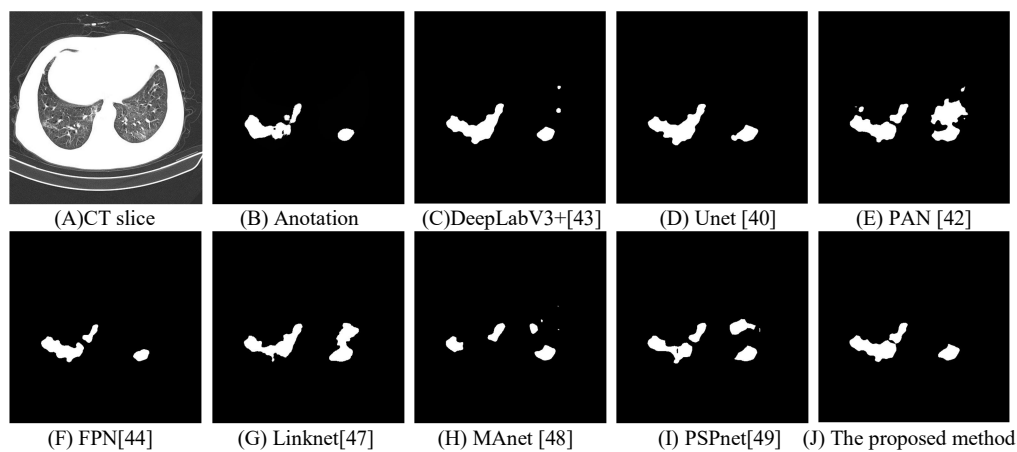

**FIGURE 9|** Segmentation of COVID-19 lesions with different deep learning methods.

For visual inspection, we selected a typical CT slice from a public dataset for demonstration. The CT slice, the corresponding annotation, DeepLabV3+ [43], Unet [40], PAN [42], FPN [44], Linknet [47], MAnet [48], PSPnet [49] and the proposed method are displayed in **Figure 9**. In which, many methods [40, 43, 48, 49] used the local features to segment COVID-19 lesions, the approaches may cause parts of small lesions to be undetected. While parts of methods [42, 44, 47] exploited the local and global features for COVID-19 lesion segmentation, the strategy may cause parts of clutters to be unremoved. On the contrary, a deep-supervised ensemble learning network is presented to combine with the advantages of different deep learning networks[40, 42-44] for COVID-19 lesion segmentation. As observed, the proposed method has largely improved the segmentation results compared with seven deep learning networks [40,42-44,47-49]. In other words, by using the deep-supervised ensemble learning network, the proposed method has a <span style="color:blue">good</span> performance in COVID-19 lesion segmentation.

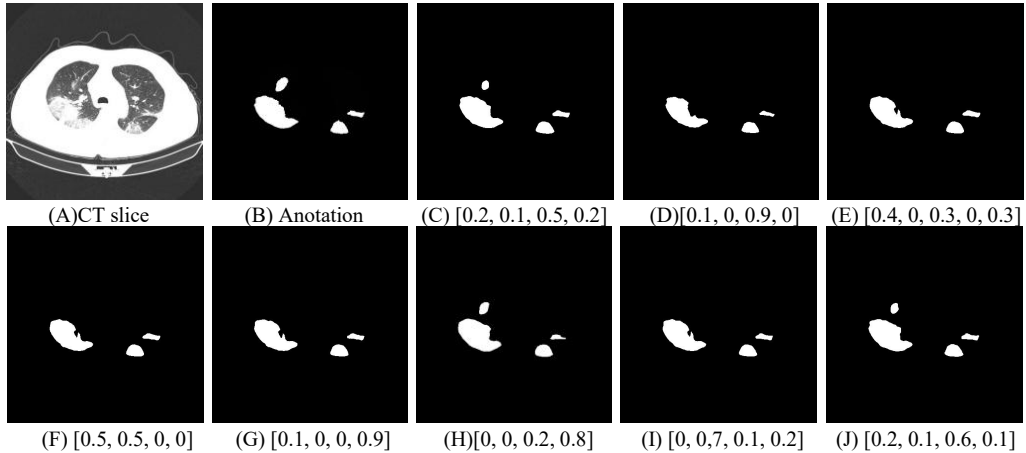

(A)CT slice (B) Anotation (C) [0.2, 0.1, 0.5, 0.2] (D)[0.1, 0, 0.9, 0] (E) [0.4, 0, 0.3, 0, 0.3]

(F) [0.5, 0.5, 0, 0] (G) [0.1, 0, 0, 0.9] (H)[0, 0, 0.2, 0.8] (I) [0, 0,7, 0.1, 0.2] (J) [0.2, 0.1, 0.6, 0.1]

**FIGURE 10|** COVID-19 lesion segmentation with different weighting parameters.

In order to accurately segment novel coronavirus pneumonia lesions, the enumeration method is applied to estimate the best weighting parameters $w_1$, $w_2$, $w_3$ and $w_4$ . In **Figure 10**, COVID-19 lesion segmentation results with different weighting parameters are displayed to evaluate the validation of the proposed method. It can be seen that the best weighting papameters are 0.2, 0.1, 0.6 and 0.1.

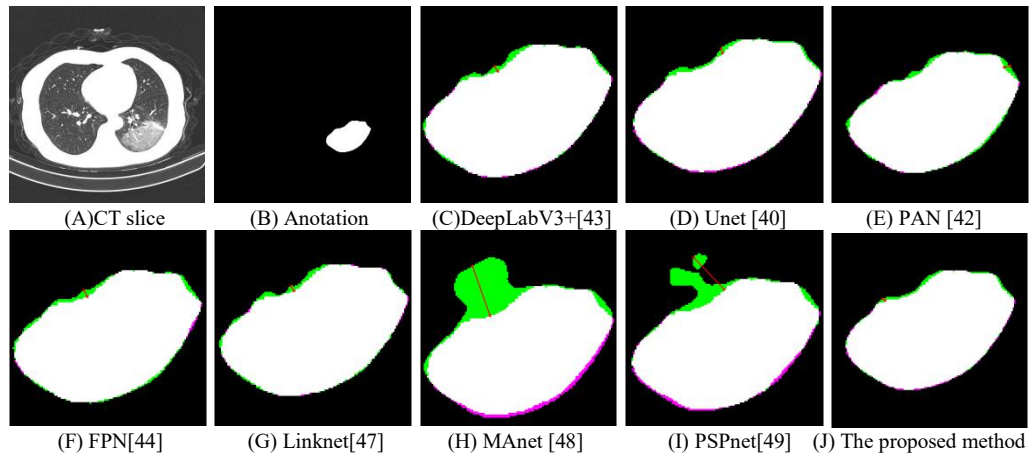

(A)CT slice (B) Anotation (C)DeepLabV3+[43] (D) Unet [40] (E) PAN [42]

(F) FPN[44] (G) Linknet[47] (H) MAnet [48] (I) PSPnet[49] (J) The proposed method

**FIGURE 11|** Hausdorff distance with different methods.

<span style="color:blue">To investigate the effect of the proposed method in COVID-19 lesion segmentation, a CT slice is chosen in **Figure 11(A)** and the corresponding annotation data is shown in **Figure 11(B)**. In order to better illustrate the effectiveness of the proposed method, the detected COVID-19 lesions with different methods were magnified. As shown in **Figure 11**, the pink region is the annotations, the green region is the predictions, and the white area is the common region covered by both predictions and annotations. In **Figure 11(C-J)**, the hausdorff distance with different methods are 3.6056, 3.1623, 4.1231, 4.4721, 3.6056, 32.8938, 27.5862 and 2.2361. In other words, the proposed method has a lower hausdorff distance than these typical methods [40,42-44,47-49].</span>

### 3.2 Quantitative evaluation

We employed a dataset from the China consortium of chest CT image investigation, the dataset include 750 CT images. As observed in **Table IV**, The IoU values corresponding to the DeepLabV3+ [43], Unet [40], PAN [42], FPN [44], Linknet [47], MAnet [48], PSPnet [49] and the proposed method are 0.7058, 0.6927, 0.7031, 0.7081, 0.6883, 0.6067, 0.6696 and 0.7279. Whereas the $F_1$ values corresponding to different methods are 0.7886, 0.7720, 0.7881, 0.7931, 0.7618, 0.7216, 0.7557 and 0.8065, and the hausdorff distance with different methods are 99.8686, 96.9404, 97.6020, 98.9161, 101.9168, 112.6191, 104.5176 and 92.4604. Both visual inspection and quantitative evaluation exhibited that our method can outperform these typical methods[40,42-44,47-49].

Table IV The IoU values with different methods

| Method | IoU | $F_1$ | H |
|---|---|---|---|
| DeepLabV3+[43] | 0.7058 | 0.7886 | 99.8686 |
| Unet[40] | 0.6927 | 0.7720 | 96.9404 |
| PAN[42] | 0.7081 | 0.7931 | 97.6020 |
| FPN[44] | 0.7031 | 0.7881 | 98.9161 |
| Linknet[47] | 0.6883 | 0.7618 | 101.9168 |
| MAnet[48] | 0.6067 | 0.7216 | 112.6191 |
| PSPnet[49] | 0.6696 | 0.7557 | 104.5176 |
| The proposed method | **0.7279** | **0.8065** | **92.4604** |

The relationship between the weighting parameters and the IoU index is shown in **Table V**. Both visual inspection and quantitative evaluation exhibited that the best weighting papameters are 0.2, 0.1, 0.6 and 0.1.

Table V The relationship between the weighting parameters and the IoU index.

| $w_1$ | $w_2$ | $w_3$ | $w_4$ | IoU |
|---|---|---|---|---|
| 0.1 | 0.0 | 0.9 | 0.0 | 0.6929 |
| 0.4 | 0.3 | 0.0 | 0.3 | 0.7083 |
| 0.0 | 0.7 | 0.1 | 0.2 | 0.7101 |
| 0.4 | 0.3 | 0.1 | 0.2 | 0.7108 |
| 0.7 | 0.1 | 0.1 | 0.1 | 0.7116 |
| 0.1 | 0.1 | 0.2 | 0.6 | 0.7232 |
| 0.2 | 0.3 | 0.4 | 0.1 | 0.7219 |
| 0.3 | 0.0 | 0.1 | 0.6 | 0.7150 |
| 0.5 | 0.1 | 0.2 | 0.2 | 0.7142 |
| 0.6 | 0.0 | 0.2 | 0.2 | 0.7120 |
| 0.8 | 0.0 | 0.0 | 0.2 | 0.7075 |
| 0.0 | 0.1 | 0.8 | 0.1 | 0.6951 |
| 0.2 | 0.0 | 0.7 | 0.1 | 0.7252 |
| 0.0 | 0.1 | 0.6 | 0.3 | 0.7229 |
| 0.2 | 0.1 | 0.6 | 0.1 | **0.7279** |

## 4    DISCUSSION

In this paper, a deep-supervised ensemble learning network is presented for COVID-19 segmentation in CT images, the proposed method has many specific characteristics and advantages. Based on the fact that a large number of COVID-19 CT images and the corresponding lesion annotations are difficult to obtained, a transfer learning strategy is employed to make up for the shortcoming and alleviate the overfitting problem. Second, a unique ensemble module is presented to improve the segmentation performance. While many studies use only one neural network to segment COVID-19 lesions, which cannot effectively discriminate COVID-19 lesions and other unrelated structures in CT images. Third, the proposed method is expected to preserve the completeness of COVID-19 lesions while maximally eliminating the unrelated structures. Last, the proposed method has a good performance in COVID-19 lesion segmentation in CT images.

The proposed method was validated in a publicly available dataset from the China consortium of chest CT image investigation. Both visual inspection and quantitative evaluation exhibited that the proposed approach can outperform these typical deep learning networks in COVID-19 lesion segmentation [40,42-44,47-49]. Compared with manually defined annotations, our methods obtained a higher accuracy in COVID-19 lesion segmentation with IoU index of 0.7279 and $F_1$ value of 0.8065 than these typical methods. In which, many typical methods [40, 43, 48, 49] used the local features to segment COVID-19 lesions, the approaches may cause parts of small lesions to be undetected. While parts of methods [42, 44, 47] exploited the local and global features for COVID-19 lesion segmentation, the strategy may cause parts of clutters to be unremoved. On the contrary, a deep-supervised ensemble learning network is presented to combine with the advantages of different deep learning networks[40, 42-44] for COVID-19 lesion segmentation. In other words, the proposed method uses weight parameters to measure the importance of local and global features for COVID-19 lesion segmentation. While the compared methods used only one neural network to segment COVID-19 lesions, it may cause parts of COVID-19 lesions to be undetected.

Compared with these conventional neural networks [40,42-44,47-49], the proposed method appears more efficient on COVID-19 lesion segmentation. This is ascribed to a well-designed fusion of transfer learning strategy, data augmentation and multiple neural networks ensemble approach. In other words, the proposed method outperforms the conventional methods in that the merits of local and global features are efficiently combined. In addition, segmentation of COVID-19 lesions has important clinical research significance. It can help doctors to diagnosis COVID-19 and develop the best treatment plan.

However, the designed deep-supervised ensemble learning network may take up more time and space than tranditional conventional neural networks [40,42-44,47-49]. Additionally, based on the fact that a large number of COVID-19 CT images and the corresponding lesion annotations are difficult to obtained, accurate segmentation of small coronavirus pneumonia lesions is still a long way off.

In conclusion, a deep-supervised ensemble learning network is presented for coronavirus pneumonia lesion segmentation in CT images. Based on the reality that mounts of COVID-19 CT images and the corresponding lesion annotations are difficult to acquired, a transfer learning strategy is used to alleviate the overfitting problem in small dataset. Another contribution of the proposed method concerns on the deep-supervised ensemble learning network. Using a single deep learning network, the accuracy of COVID-19 lesion segmentation results can't reach a satisfactory performance. To overcome the problem, an ensemble strategy is presented to integrate multiple deep learning networks for COVID-19 lesion and its boundary segmentation in CT images. Experimental results indicated that our proposed deep-supervised ensemble learning model has a good performance in COVID-19 lesion and its boundary segmentation in CT images.

## DATA AVAILABILITY STATEMENT

The datasets applied in this paper can be aquired in the website: http://ncov-ai.big.ac.cn/download. Which is a publicly available dataset.

## ETHICS STATEMENT

Ethical review and approval was not required for the study on human participants in accordance to the local legislation and institutional requirements.

## AUTHOR CONTRIBUTIONS

YP was major contributors in writing the manuscript. All authors analyzed the data.

## FUNDINGS

This research was funded by the Jiangxi Provincial Natural Science Foundation (Nos. 20212BAB202007, 20202BAB212004, 20204BCJL23035, 20192ACB21004, 20181BAB202017), by the Educational Science Research Project of China Institute of communications Education (No. JTYB20-33), by the Huxiang high-level talent gathering project (No. 2019RS1072), by the Scientific and Technological Research Project of Education Department in Jiangxi Province (No. GJJ190356) and the Science and Technology project of Changsha City (No. kq2001014).

## ACKNOWLEDGMENTS

We are grateful to the China consortium of chest CT image investigation for COVID-19 data sharing.

## SUPPLEMENTARY MATERIAL

The Supplementary Material for the article can be found online at : https://www. frontiersion.org.

## REFERENCES

1. Alon D, Paitan Y, Robinson E, Ganor N, Lipovetsky J, Yerushalmi R, et al. Downregulation of CD45 Signaling in COVID-19 patients is reversed by C24D, a novel CD45 targeting peptide. Front. Med. (2021) 8:1251. doi: 10.3389/fmed.2021.675963

2. Murphy A, Pinkerton LM, Bruckner E, Risser HJ. The impact of the novel coronavirus disease 2019 on therapy service delivery for children with disabilities. The Journal of Pediatrics (2021) 231:168-177. doi: 10.1016/j.jpeds.2020.12.060

3. Comunale BA, Engineer L, Jiang Y, Andrews JC, Liu Q, Ji L, et al. Poliovirus vaccination induces a humoral immune response that cross reacts with SARS-CoV-2. Front. Med. (2021) 8:1285. doi: 10.3389/fmed.2021.710010

4. Li L, Cao Y, Fan J, Li T, Lang J, Zhang H, et al. Impact of COVID-19 pandemic on the clinical activities in obstetrics and gynecology: a national survey in China. Front. Med. (2021) 8:1225. doi: 10.3389/fmed.2021.633477

5. Meng Z, Wang T, Chen L, Chen X, Li L, Qin X, et al. The effect of recombinant human interferon alpha nasal drops to prevent COVID-19 pneumonia for medical staff in an epidemic area. Curr. Top. Med. Chem. (2021) 21:920-927. doi: 10.2174/1568026621666210429083050

6. Herpe G, Lederlin M, Naudin M, Obana M, Chaumoitre K, Gregory J, et al. Efficacy of chest CT for COVID-19 pneumonia diagnosis in France. Radiology (2021) 298:81-87. doi: 10.1148/radiol.2020202568

7. Zhao Y, Ji D, Li Y, Zhao X, Lv W, Xin X, et al. Three-dimensional visualization of microvasculature from few-projection data using a novel CT reconstruction algorithm for propagation-based X-ray phase-contrast imaging. Biomed. Opt. Express (2020) 11:364-387. doi: 10.1364/BOE.380084

8. Schalekamp S, Bleeker-Rovers CP, Beenen LFM, Quarles van Ufford HME. Gietema, HA, Stoger JL, et al. Chest CT in the emergency department for diagnosis of COVID-19 pneumonia: dutch experience. Radiology (2021) 298:98-106. doi: 10.1148/radiol.2020203465

9. Bhargava A, Bansal A. Novel coronavirus (COVID-19) diagnosis using computer vision and artificial intelligence techniques: a review. Multimed. Tools Appl. (2021) 80:19931-19946. doi: 10.1007/s11042-021-10714-5

10. Peng Y, and Xiao C. An oriented derivative of stick filter and post-processing segmentation algorithms for pulmonary fissure detection in CT images. Biomed. Signal Proces. (2018) 43:278-288. doi: 10.1016/j.bspc.2018.03.013

11. Peng Y, Zhong H, Xu Z, Tu H, Li X, Peng L. Pulmonary lobe segmentation in CT images based on lung anatomy knowledge. Math. Probl. Eng. (2021) 2021:5588629. doi: 10.1155/2021/5588629

12. Zhang Y, Hu Y, Zhao S, Cui C. The utility of PET/CT metabolic parameters measured based on fixed percentage threshold of SUVmax and adaptive iterative algorithm in the new revised FIGO staging system for stage III cervical cancer. Front. Med. (2021) 8:1189. doi: 10.3389/fmed.2021.680072

13. Chen S, Chen J, Yang Y, Chen C, Wang M, Lin L. Use of radiographic features in COVID-19 diagnosis: challenges and perspectives. Journal of Chinese Medical Association (2020) 83:644-647. doi: 10.1097/JCMA.0000000000000336

14. Dastidar TR, Ethirajan R. Whole slide imaging system using deep learning-based automated focusing. Biomed. Opt. Express (2020) 11:480-491. doi: 10.1364/BOE.379780

15. Ali MJ, Hanif M, Haider MA, Ahmed MU, Sundas F, Hirani A, et al. Treatment options for COVID-19: a review. Front. Med. (2020) 7:480. doi: 10.3389/fmed.2020.00480

16. Hammoudi K, Benhabiles H, Melkemi M, Dornaika F, Arganda-Carreras I, Collard D, et al. Deep learning on chest X-ray images to detect and evaluate pneumonia cases at the ear of COVID-19. J. Med. Syst. (2021) 45:75. doi:10.1007/s10916-021-01745-4

17. Erion G, Janizek JD, Sturmfels P, Lundberg SM, Lee S, Improving performance of deep learning models with axiomatic attribution priors and expected gradients. Nature Machine Intelligence. (2021) 37:1-12. doi: 10.1038/s42256-021-00343-w

18. Yazdekhasty P, Zindar A, Nabizadeh-ShahreBabak Z, Roshandel R, Khadivi P, Karimi N, et al. Bifurcated autoencoder for segmentation of COVID-19 infected regions in CT images. arXiv preprint 2011. (2020) 00631.

19. Wang Y, Zhang Y, Liu Y, Tian J, Zhong C, Shi Z, et al. Does non-COVID-19 lung lesion help? Investigating transferability in COVID-19 CT image segmentation. Comput. Meth. Prog. Bio. (2021) 202:106004. doi: 10.1016/j.cmpb.2021.106004

20. Gao K, Su J, Jiang Z, Zeng L, Feng Z, Shen H, et al. Dual-branch combination network (DCN): Towards accurate diagnosis and lesion segmentation of COVID-19 using CT images. Med. Image. Anal. (2021) 67:101836. doi: 10.1016/j.media.2020.101836

21. Zhao S, Li Z, Chen Y, Zhao W, Xie X, Liu J, et al. SCOAT-Net: A novel network for segmentation COVID-19 lung opacification from CT images. Pattern. Recogn. (2021) 119:108109. doi: 10.1016/j.patcog.2021.108109

22. Abdel-Basset M, Chang V, Hawash H, Chakrabortty RK, Ryan M. FSS-2019-nCov: a deep learning architecture for semi-supervised few-shot segmentation of COVID-19 infection. Knowl-Based. Syst. (2021) 212:106647. doi: 10.1016/j.knosys.2020.106647

23. Yang D, Xu Z, Li W, Myronenko A, Roth HR, Harmon S, et al. Federated semi-supervised learning for COVID region segmentation in Chest CT using multi-national data from China, Italy, Japan. Med. Image. Anal. (2021) 70:101992. doi: 10.1016/j.media.2021.101992

24. Piccolo V, Neri I, Filippeschi C, Oranges T, Argenziano G, Battarra VC, et al. Chilblain-like during COVID-19 epidemic: a preliminary study on 63 patients. J. Eur. Acad. Dermatol. (2020) 34. doi: 10.1111/jdv.16526

25. Yang Y, Chen J, Wang R, Ma T, Wang L, Chen J, et al. Towards unbiased COVID-19 lesion localisation and segmentation via weakly supervised learning. IEEE 18th International Symposium on Biomedical Imaging. (2021) 1966-1970. doi: 10.1109/ISBI48211.2021.9433806

26. Laradji I, Rodriguez P, Manas O, Lensink K, Law M, Kurzman L, et al. A weakly supervised consistency-based learning method for COVID-19 segmentation in CT images. IEEE Winter Conference on Applications of Computer Vision. (2021) 2453-2462.

27. Wu X, Chen C, Zhong M, Wang J, Shi J. COVID-AL: the diagnosis of COVID-19 with deep active learning. Med. Image. Anal. (2021) 68:101913. doi: 10.1016/j.media.2020.101913

28. Wang X, Deng X, Fu Q, Zhou Q, Feng J, Ma H, et al. A weakly supervised framework for COVID-19 classification and lesion localization from chest CT. IEEE. Trans. Med. Imaging (2020) 39:2615-2625. doi: 10.1109/TMI.2020.2995965

29. Yao Q, Xiao L, Liu P, Zhou SK. Label-free segmentation of COVID-19 lesions in lung CT. IEEE. Trans. Med. Imaging. (2021). doi: 10.1109/TMI.2021.3066161

30. Zhang K, Liu X, Shen J, Li Z, Sang Ye, Wu X, et al.Clinically applicable AI system for accurate diagnosis, quantitative measurements, and prognosis of COVID-19 pneumonia using computed tomography. Cell (2020) 181:1423-1433. doi: 10.1016/j.cell.2020.04.045

31. Zhu Y, Yeung CH, Lam EY. Digital holographic imaging and classification of microplastics using deep transfer learning. Appl. Optics. (2021) 60:38-47. doi: 10.1364/AO.403366

32. Fu Y, Lei Y, Wang T, Curran WJ, Liu T, Yang X. Deep learning in medical image registration: a review. Phys. Med. Biol. (2020) 65:20TR01. doi: 10.1088/1361-6560/ab843e

33. Du S, Du J, Tang Y, Ouyang H, Tao Z, Jiang T. Achieving efficient inverse design of low-dimensional heterostructures based on a vigorous scalable multi-task learning network. Opt. Express. (2021) 29:19727-19742. doi: 10.1364/OE.426968

34. Christensen CN, Ward EN, Lu M, Lio P, Kaminski CF. ML-SIM: universal reconstruction of structured illumination microscopy images using transfer learning. Biomed. Opt. Express. (2021) 12:2720-2733. doi: 10.1364/BOE.414680

35. Wang SH, Fernandes S, Zhu Z, Zhang YD. AVNC: attention-based VGG-style network for COVID-19 diagnosis by CBAM. IEEE. Sens. J. (2021) doi: JSEN.2021.3062442

36. Krizhevsky A, Sutskever I, Hinton GE. ImageNet classification with deep convolutional neural networks. Commun. Acm. (2017) 60:84-90. doi: 10.1145/3065386

37. Marques G, Agarwal D, Diez IDLT. Automated medical diagnosis of COVID-19 through EfficientNet convolutional neural network. Appl. Soft. Comput. (2020) 96:106691. doi: 10.1016/j.asoc.2020.106691

38. Munien C, Viriri S. Classification of hematoxylin and eosinstained breast cancer histology microcopy images using transfer learning with EfficientNets. Comput. Intel. Neurosc. (2021) 2021. doi: 10.1155/2021/5580914

39. Xie Q, Luong M, Hovy E, Le QV. Self-training with noisy student improves ImageNet classification. CVPR. (2020)

40. Su R, Zhang D, Liu J, Cheng C. MSU-Net: Multi-scale U-Net for 2D medical image segmentation. Front. Genet. (2021) 12. doi: 10.3389/fgene.2021.639930

41. Wang B, Yang J, Ai J, Luo N, An L, Feng H, Yang Zu. Accurate tumor segmentation via octave convolution neural network. Frontiers in Medicine (2021) 8:653913. doi: 10.3389/fmed.2021.653913

42. Li H, Xiong P, An J, Wang L. Pyramid attention network for semantic segmentation. arXiv preprint (2018) 1805.

43. Chen L C, Zhu Y, Papandreou G, Schroff F, Adam H. Encoder-decoder with atrous separable convolution for semantic image segmentation. ECCV, (2018) 801-818.

44. Lin TY, Dollar P, Girshick R, He K, Hariharan B, Belongie S. Feature pyramid networks for object detection. CVPR, (2017) 2117-2125.

45. Golla AK, Bauer DF, Schmidt R, Russ T, Norenberg D, Chung K, et al. Convolutional neural network ensemble segmentation with ratio-based sampling for the arteries and veins in abdominal CT scans. IEEE. Trans. Bio-med. Eng. (2021) 68:1518-1526. doi: 10.1109/TBME.2020.3042640

46. Hartley MG, Ralph E, Norville IH, Prior JL, Atkins TP. Comparison of PCR and viable count as a method for enumeration of bacteria in an A/J mouse aerosol model of Q fever. Front. Microbiol. (2019) 10:1552. doi: 10.3389/fmicb.2019.01552

47. Chaurasia A, Culurciello E. Linknet: exploiting encoder representations for efficient semantic segmentation. VCIP (2017) 1-4. doi: 10.1109/VCIP.2017.8305148

48. Fan T, Wang G, Li Y, Wang H. MA-Net: A Multi-Scale attention network for liver and tumor segmentation. IEEE Access. (2020) 8:179656-179665. doi: 10.1109/ACCESS.2020.3025372

49. Zhao H, Shi J, Qi X, Wang X, Jia J. Pyramid scene parsing network. CVPR. (2017) 2881-2890.

**Conflict of Interest:** The authors have no conflicts with others.